\documentclass[preprint,12pt]{elsarticle}

\usepackage{color}
\usepackage{amssymb}

\journal{Solid State Communications}

\begin{document}

\begin{frontmatter}



\title{Indirect band gap in graphene from modulation of the  Fermi velocity}

\author{Jonas R. F. Lima} 
\address{ Instituto de Ciencia de Materiales de Madrid (CSIC) - Cantoblanco, Madrid 28049, Spain}
\ead{jonas.iasd@gmail.com}
\author{  F. Moraes}

\address{Departamento de F\'{\i}sica, CCEN,  Universidade Federal 
da Para\'{\i}ba, Caixa Postal 5008, 58051-970 , Jo\~ao Pessoa, PB, Brazil}

\begin{abstract}
In this work we study theoretically the electronic properties of a sheet of graphene grown on a periodic heterostructure substrate. We write an effective Dirac equation, which includes a dependence of both the band gap and the Fermi velocity on the position, due to the influence of the substrate. This way, both  bandgap and  Fermi velocity enter the Dirac equation as operators.   The Dirac equation is solved exactly and  we find the superlattice minibands with gaps due to the breaking of translational symmetry induced by the underlying heterostructure. The spatial dependence of the Fermi velocity makes the band gap be indirect, bringing about interesting possibilities for applications in the design of nanoelectronic devices. In the limit of constant Fermi velocity we obtain a  band structure, with direct band gap, very close to the one previously found in the literature, obtained  using the transfer matrix method. 
\end{abstract}

\begin{keyword}



\end{keyword}

\end{frontmatter}


\section{Introduction}

Most semiconductor devices presently being developed and produced have two or more kinds of semiconductor materials composing, what is called a  heterostructure \cite{dutta2003advanced}. The variation of the composition is used to control fundamental parameters in semiconductor crystals and devices, such as the motion of electrons and holes, the effective mass of charge carriers, and the electron energy spectrum, through band engineering. One important feature of heterostructures is the possibility of restricting the movement of charge carriers, creating quantum confinement. Examples of confining heterostructures  are quantum dots, quantum wires and quantum wells. Transport properties of  a semiconductor  may be obtained using the effective mass theory \cite{Gregory,Slaterpr,Jamespr}. In this approach, electrons (or holes) are described as free particles with an effective mass that incorporates the interaction between the carriers and the lattice structure. Since the effective mass of the carriers depends on the material, in a heterostructure it depends on their position. The effective mass is then an operator which does not commute with momentum implying that the usual kinetic operator $p^2/2m(x)$ is not Hermitian and have to be modified. Effective mass operators have been largely used to obtain the minibands associated with periodicity of  heterostructures \cite{jrflima,gadella,Gadella2007265,gadella2,einevoll,morrow,PhysRevB.42.3485}. 

Since its first successful experimental fabrication in 2004 \cite{Novoselov}, graphene has attracted a great deal of attention due to its interesting properties and the wide range of application. Graphene is an atomically thin layer made out of carbon atoms arranged in a hexagonal structure which presents an extremely efficient electronic conductivity with an easy control, which makes it a material with a great application potential for the fabrication of electronic devices. However, the physics of low-energy carriers in graphene is described by an effective Dirac Hamiltonian. There is no gap in the band structure of the ideal graphene sheet, so the Dirac electron becomes massless and cannot be confined by an electrostatic potential due to the Klein tunneling \cite{Katsnelson}, which impedes its use in electronic devices. Nevertheless, it was shown that it is possible to suppress the Klein tunneling with a magnetic field \cite{Roy,Maksym} and with a spatially modulated gap \cite{peres,Giavaras}, leading to confined states. For a recent review on the various possible ways to avoid Klein tunneling in graphene see reference \cite{jaku}.

A periodic modulation of the band gap of graphene leads to graphene-based superlattices. Recently, graphene has been successfully used in heterostructures for making atomically thin circuitry \cite{Levendorf}, high-responsivity photodetectors \cite{Xiaomu} and field-effect transistors \cite{Moon}, among other high-technology applications. Superlattice structures based on graphene have been studied in many forms: modulation of the gap by an externally applied periodic potential \cite{PhysRevLett.101.126804}, alternating graphene and h-BN nanoribbons \cite{PhysRevB.78.245402}, alternating graphene and graphane nanoribbons \cite{PhysRevB.84.113413}, among others. In this work we study  a heterostructure solely made of graphene, without doping and without material discontinuities. This is done by taking advantage of the influence that the substrate has on graphene. Our aim is to study the combined effects of spatial modulation of both the gap and the Fermi velocity. For this, we choose the substrate studied in reference \cite{ratnikov} which considered the bandgap variation but not the Fermi velocity dependence on the position.

It is very well know that the electronic structure of graphene is affected by its substrate. For instance,  a hexagonal boron nitride (\textit{h}-BN) substrate leads to the opening of a gap of the order of $30$ meV  which depends on the commensurability between the lattices \cite{guinea1,guinea2},  while  SiC   induces  a gap of $\approx0.26$ eV in graphene \cite{SiC}. Graphene deposited on a SiO$_{2}$ surface has a gap of $\approx0.35$ eV for an oxygen-terminated surface, whereas there is no band gap when the oxygen atoms are passivated with hydrogen atoms \cite{Philip}. It is also on the SiO$_{2}$ substrate where graphene shows the lowest quality properts, exhibiting characteristics that are very inferior to the expected in  intrinsic graphene \cite{Ishigami,Fratini,Dean}. Recently, it was pointed out that graphene on \textit{h}-BN substrate probably elastically deforms itself in order to reduce the lattice mismatch between the two structures \cite{Eckmann}. This strain induced by the substrate has important consequences on the electronic properts of graphene \cite{Mauricio}. Another important  electronic property of graphene that depends on the substrate is the Fermi velocity. For graphene on different substrates the Fermi velocity has been measured by different authors and their results  summarized in \cite{Hwang}. 

Taking advantage of the influence of the substrate on the electronic properties of graphene, P. V. Ratnikov \cite{ratnikov} in 2009 proposed a method to spatially modulate the band gap of graphene by considering a pristine graphene sheet on a heterostructured substrate composed by alternating \textit{h}-BN and SiO$_{2}$ strips. Developments of this idea have appeared in the recent literature \cite{PhysRevB.86.205422,Li20132895,Kryuchkov2013524,JAP:9221137} with and without a modulated externally applied field. In all these papers it was considered that the heterostructured substrate induces only an energy gap modulation, and the remaining effects of the substrate on graphene were not taken into account. However, this description was able to elucidate the effects of the energy gap modulation on the electronic structure of graphene. One of the effects neglected in these papers is the variation of the Fermi velocity. The importance of the Fermi velocity variation has been pointed out in the literature \cite{PhysRevB.86.205422} but only in two cases incorporated in the study of graphene-based superlattices:  the case of a bilayer \cite{hosein} and the case of scattering across a heterostructure in the monolayer \cite{peres}. To the best of our knowledge, ours is the first work to present a study of the band structure of the graphene superlattice induced by the sustrate, including both modulation of the band gap and Fermi velocity. We will see below that this dependence of the Fermi velocity on the substrate leads to a direct to indirect band gap transition in graphene. We also find that the shape of the minibands is quite sensitive to the inclusion  of the spatial variation of the Fermi velocity. In our study, for the sake of comparison, we use the same parameter values as in \cite{ratnikov}, including the \textit{h}-BN induced bandgap of 53 meV in graphene even though more recent works  \cite{guinea1,guinea2} indicate not only a lower value but also a dependence on the relative position between the graphene and the \textit{h}-BN lattices.

In this work we consider a graphene sheet deposited in a heterostructured substrate composed of two different materials. We consider that in the graphene layer  there  will be a jump in the Fermi velocity  and in the  band gap at the positions corresponding to the  junctions between the materials composing the substrate. We use an effective Dirac equation with position dependent gap term (mass) and Fermi velocity, which describes the low-energy states in graphene. By exactly solving the Dirac equation we find the superlattice minibands induced by the substrate heterostructure. We compare our results with the previous one \cite{ratnikov}, where the Dirac equation with a periodically modulated band gap was solved using the transfer matrix method. In our work we include the spatial variation of the Fermi velocity, which will induce an indirect band gap transition in graphene. An indirect band gap has been realized  in graphene nanoribbons \cite{majumdar}, in graphene bylayers under field modulation \cite{raza}, in a graphene-graphane superlattice \cite{PhysRevB.84.113413}  and in a single-layer graphene sheet with combined electric and magnetic fields \cite{xin}.  To the best of our knowledge, what we present here is the first case of Fermi velocity modulation yielding an indirect gap for a pristine single layer of graphene.  Modulation of the Fermi velocity for a graphene bilayer was done  in reference\cite{hosein}, which also lead to indirect gaps. We remark that there are other important effects of the substrate that we are not considering here as, for example, the effect of the strain on the graphene lattice caused by the substrate. However, this is justified since the aim of this work is to elucidate the effect of a variable Fermi velocity on graphene and compare two methods of calculating the dispersion relation. The incorporation of further substract influence on graphene will enrich the model and will be the subject of a future publication.

The paper is organized as follows. In Section II we write out the effective Dirac equation for low energy electronic states in graphene with a variable mass term, which means the energy gap, and Fermi velocity, taking into account that the Hamiltonian has to be Hermitian. In Section III we use the effective Dirac equation to find the superlattice miniband structure of graphene on the heterostructured substrate. The paper is summarized  in the conclusions.

\section{Effective Dirac Equation with variable gap term and Fermi velocity}

We start by generalizing the two-dimensional effective Dirac equation for low-energy states in graphene for the case where the Fermi velocity and gap term depend on the position. Naively, we write 
\begin{equation}
\bigl[-i\hbar v_F(x)(\sigma_x \partial_x +\sigma_y \partial_y) + \Delta(x)\sigma_z\bigr]\psi = E \psi , \label{dirac1}
\end{equation}
where $\sigma_i$ are the Pauli matrices acting in the two graphene sublattices, $v_F(x)$ is the Fermi velocity, $\Delta(x)$ is the gap term and the spinor
\begin{equation}
\psi=
\left( 
\begin{array}{c}
\psi_A(x,y)\\
\psi_B(x,y)
\end{array}
\right)
\label{psi} ,
\end{equation}
with $A$ and $B$ representing the two polarizations of the \textit{pseudospin} that correspond to the graphene sublattices.

Since the Fermi velocity and the gap term both depend on position, they are operators. Furthermore, they do not commute with the linear momentum, which make the Hamiltonian non-Hermitian. To fix this, a generalization of the Dirac Hamiltonian with spatial dependent Fermi velocity was obtained from an effective low-energy theory of a tight binding model in \cite{peres}, which is equivalent to consider the permutations between the operators in the first term of the Dirac operator. This way, Eq. (\ref{dirac1}) becomes

\begin{eqnarray}
-i\hbar \left(\frac{\sigma_x}{2} [v_F(x)\partial_x +\partial_x v_F(x)]  + \sigma_y v_F(x)\partial_y \right)\psi
 + \Delta(x)\sigma_z \psi = E \psi .
\label{diraceq}
\end{eqnarray}
Notice that for $v_F = const.$ we recover Eq. (1) of reference \cite{ratnikov} with the potential $V=0$.

Writing 
\begin{equation}
\psi = e^{-ik_y y}\left(
\begin{array}{c}
\psi_A(x) \\
\psi_B(x)
\end{array}\right)
\end{equation}
and replacing it in (\ref{diraceq}), one gets two coupled equations 

\begin{eqnarray}
-\frac{i \hbar}{2} \left[v_F(x)\frac{\partial}{\partial_x}+\frac{\partial}{\partial_x}v_F(x)\right]\psi_B + i\hbar v_F(x)k_y \psi_B 
=(E- \Delta(x))\psi_A
\end{eqnarray}
and
\begin{eqnarray}
-\frac{i \hbar}{2} \left[v_F(x)\frac{\partial}{\partial_x}+\frac{\partial}{\partial_x}v_F(x)\right]\psi_A -i\hbar v_F(x)k_y \psi_A 
=(E+\Delta(x)) \psi_B .
\end{eqnarray}

Uncoupling these equations one gets

\begin{eqnarray}
-\frac{\hbar^2}{4} \left\{ v_F\frac{\partial}{\partial x}\left[\frac{1}{E- \Delta}\left( v_F\frac{\partial}{\partial x}+\frac{\partial}{\partial x} v_F\right)
 -\frac{2v_F k_y}{E- \Delta}\right] +\frac{\partial}{\partial x}v_F\left[\frac{1}{E- \Delta}\left( v_F\frac{\partial}{\partial x}\right.\right. \right. \nonumber \\
\left.\left. \left.  + \frac{\partial}{\partial x}v_F\right) -\frac{2v_F k_y}{E- \Delta}\right]+\left[\frac{2v_F k_y}{E- \Delta}\left(v_F\frac{\partial}{\partial x}+\frac{\partial}{\partial x}v_F\right) 
 -\frac{4v_F^2k_y^2}{E- \Delta} \right] \right\}\psi_B = (E+ \Delta)\psi_B 
\label{uncoupled}
\end{eqnarray}

and 

\begin{eqnarray}
-\frac{\hbar^2}{4} \left\{ v_F\frac{\partial}{\partial x}\left[\frac{1}{E+ \Delta}\left(v_F\frac{\partial}{\partial x}+\frac{\partial}{\partial x}v_F\right)
 +\frac{2v_F k_y}{E+ \Delta}\right] +\frac{\partial}{\partial x}v_F\left[\frac{1}{E+ \Delta}\left(v_F\frac{\partial}{\partial x}\right. \right. \right. \nonumber \\
\left. \left. \left. +\frac{\partial}{\partial x}v_F\right) +\frac{2v_F k_y}{E+ \Delta}\right]-\left[\frac{2v_F k_y}{E+ \Delta}\left(v_F\frac{\partial}{\partial x}+\frac{\partial}{\partial x}v_F\right)
 +\frac{4v_F^2k_y^2}{E+ \Delta}\right] \right\}\psi_A= (E-\Delta)\psi_A
\label{uncoupled2}
\end{eqnarray}

Before going to  the next section, where we will use the uncoupled Eqs.(\ref{uncoupled}) and (\ref{uncoupled2}) to find the band structure of the graphene superlattice, we take a look at their symmetries. It is easily seen that, by  simultaneously making $E \rightarrow -E$ and $k_y \rightarrow -k_y$ the Eqs. (\ref{uncoupled}) and (\ref{uncoupled2}) are transformed into each other. As it will be seen below this has important consequences in the type (direct or indirect) of the band gap.

\section{The Superlattice Minibands}

In this section, we use the effective Dirac Hamiltonian obtained in the last section to describe electrons in a graphene layer on a periodic heterostructure substrate. In an ordinary semiconductor heterostructure the charge carriers may be described as free Schr\"odinger  electrons (or holes) with an effective mass \cite{Gregory,Slaterpr,Jamespr} that depends on the material and, therefore, on position. This approach has been largely used to obtain the minibands associated with periodicity of the heterostructure (see for example \cite{jrflima} and references therein). In our case, we have an effective  Dirac operator with both gap term and Fermi velocity depending on position to account for the substrate influence on the graphene layer. 

\begin{figure}[h]
\centering
\includegraphics[width=10cm,height=3.5cm]{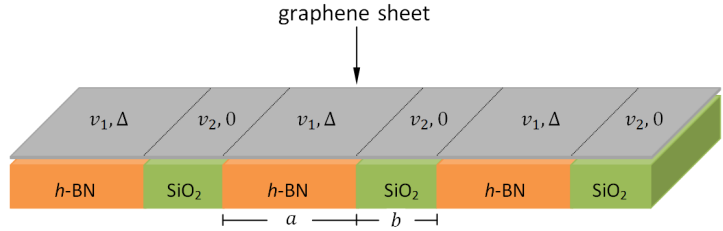}	
\caption{Graphene sheet on the periodic heterostructure substrate. Each color represents a different substrate which will induce  different gap and Fermi velocity in  graphene. }\label{substrate}
\end{figure}

We consider a graphene sheet deposited on a periodic heterostructure substrate composed by \textit{h}-BN and SiO$_2$, as shown in Fig. \ref{substrate}. This will induce a heterostructure in  graphene due to the dependence of both its Fermi velocity and band gap on the substrate material.

\begin{figure}[h!]
\centering
\includegraphics[width=10cm,height=5cm]{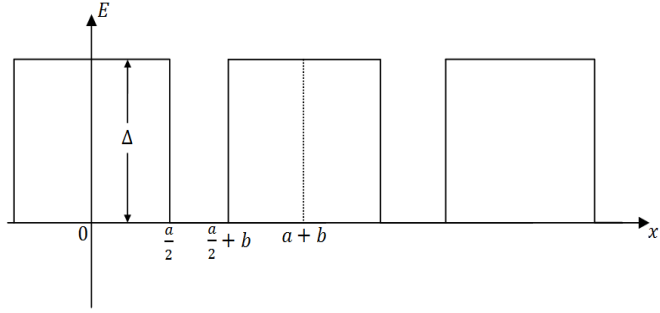}	
\caption{The periodic modulation of the bandgap in graphene due to the heterostructured substrate and the unit cell $0\leq x\leq a+b$. Graphene  has a band gap of $2\Delta = 53$ meV on \textit{h}-BN and zero gap on SiO$_2$.}\label{estrutura}
\end{figure}

As the Hamiltonian is periodic, the wave function satisfies the Bloch theorem, so

\begin{equation}
\psi(x+n(a+b), y)=e^{iKn(a+b)}\psi(x, y)\\
\label{bloch}
\end{equation}
and 
\begin{equation}
\psi'(x+n(a+b), y)=e^{iKn(a+b)}\psi'(x, y)\;,
\label{bloch1}
\end{equation}
where $K$ is the Bloch wavenumber and $n$ is an integer. Due to this fact, we can reduce our problem to the unit cell $0\leq x\leq a+b$ [see Fig.(\ref{estrutura})].

In the unit cell we have that

\begin{equation}
 \Delta (x)=\left\{ \begin{array}{cc}

                  \Delta, \ \ \ \  & 0\leq x<\frac{a}{2} ; \frac{a}{2}+b\leq x\leq a+b \\
                   0, \ \ \ \  & \frac{a}{2}\leq x<\frac{a}{2}+b \\
 \end{array}
 \right.
\end{equation}
and
\begin{equation}
 v_F (x)=\left\{ \begin{array}{cc}

                  v_1, \ \ \ \  & 0\leq x<\frac{a}{2} ; \frac{a}{2}+b\leq x\leq a+b \\
                   v_2, \ \ \ \  & \frac{a}{2}\leq x<\frac{a}{2}+b \\
 \end{array}
 \right.\;.
\end{equation}

\subsection{Boundary conditions}

\begin{figure}[hpt]
\centering
\includegraphics[width=8cm,height=3cm]{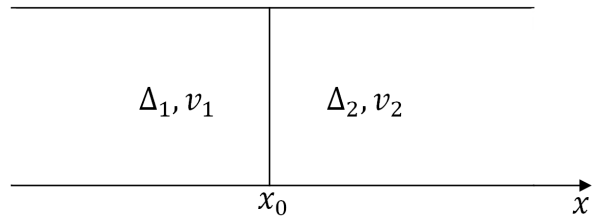}	
\caption{An interface between graphene regions with different gaps and Fermi velocities. }\label{contorno}
\end{figure}

To find the band structure it is necessary to specify the boundary conditions. We already have the Bloch conditions (\ref{bloch}) and (\ref{bloch1}), now we need the interface conditions at $x=\frac{a}{2}$ and $x=\frac{a}{2}+b$ (see Fig. 2). For this, we consider a general interface as shown in Fig. \ref{contorno}. 

In this interface, we have 
\begin{equation}
v(x)=v_1  H(x_0-x)+v_2 H(x-x_0)
\end{equation}
and
\begin{equation}
\Delta(x)=\Delta_1 H(x_0-x)+ \Delta_2 H(x-x_0),
\end{equation}
where $H$ is the Heaviside function.

Using the properties

\begin{equation}
f(x)\delta(x-x_0)=\frac{f(x_0+)+\psi(x_0-)}{2}f(x-x_0),
\end{equation}

and

\begin{eqnarray}
f(x)\delta'(x-x_0)=\frac{f(x_0+)+f(x_0-)}{2}\delta'(x-x_0) 
-\frac{f'(x_0+)+f'(x_0-)}{2}\delta (x-x_0),
\end{eqnarray}

one can write Eq. (\ref{uncoupled}) as
\begin{eqnarray}
\frac{4v_F^2}{E- \Delta}\psi_B''+\left(\frac{v_2}{E- \Delta_2}-\frac{v_1}{E-\Delta_1}\right)(v_2\psi_B'^++v_1\psi_B'^- )
\delta(x-x_0) +\frac{(v_2^2-v_1^2)}{4}\left(\frac{\psi_B^+}{E- \Delta_2}\right.\nonumber\\
\left.+\frac{\psi^-}{E- \Delta_1}\right)\delta'(x-x_0) \left(\frac{v_2^2}{E- \Delta_2}-\frac{v_1^2}{E-\Delta_1}\right)(\psi_B'^++\psi_B'^-)\delta(x-x_0) 
+\frac{(v_2-v_1)}{2}\left(\frac{v_2\psi_B^+}{E- \Delta_2}\right.\nonumber\\
\left. +\frac{v_1\psi_B^-}{E- \Delta_1}\right)\delta'(x-x_0)-2k_y\left[\left(\frac{v_2}{E-\Delta_2}-\frac{v_1}{E- \Delta_1}\right)\left(\frac{v_2\psi^+ + v_1\psi^-}{2}\right)\delta (x-x_0)\right. \nonumber \\
\left. +\left(\frac{v^2_2}{E- \Delta_2}-\frac{v_1^2}{E-\Delta_1}\right)\left(\frac{\psi^+ +\psi^-}{2}\right)\delta(x-x_0)+\frac{2v_2}{E- \Delta}\psi'\right] 
+\frac{4k_yv^2}{E- \Delta}\psi' \nonumber \\
+k_y(v_2-v_1)\left(\frac{v_2\psi^+}{E- \Delta_2}+\frac{v_1\psi^-}{E-\Delta_1}\right)\delta(x-x_0)-4\frac{k_y^2v^2}{E- \Delta}\psi=-\frac{4}{\hbar^2}(E+ \Delta)\psi_B
\label{dirac}
\end{eqnarray}

Integrating from $x_0-\epsilon$ to $x_0+\epsilon$ [see Fig.\ref{contorno}] and making $\epsilon\rightarrow 0$, one obtains
\begin{eqnarray}
\psi'^+ [4v^2_2(E- \Delta_1)+(v^2_1-v_1v_2)(E- \Delta_2)] \nonumber \\
-\psi'^- [4v^2_1(E-\Delta_2)+(v^2_2-v_1v_2)(E- \Delta_1)] \nonumber \\
+k_y[\psi^+ -\psi^-][v^2_2(E- \Delta_1)-v_1^2(E- \Delta_2)\nonumber \\
+v_1v_2(\Delta_2 -\Delta_1)]=0
\label{ic}
\end{eqnarray}

In order to obtain the boundary condition for the wave function one calculates the $primitive$ of the Eq.(\ref{dirac}), integrates it from $x_0-\epsilon$ to $x_0+\epsilon$ and take the limit  $\epsilon\rightarrow 0$. Following this procedure, we get for the boundary condition:
\begin{equation}
\psi_B^+=\beta \psi_B^- \; ,
\label{interfacepsi}
\end{equation}
where
\begin{equation}
\beta = \left(\frac{8v_1^2(E- \Delta_2)+(11v_2^2-v_1^2-2v_1v_2)(E- \Delta_1)}{8v_2^2(E- \Delta_1)+(11v_1^2-v_2^2-2v_1v_2)(E- \Delta_2)}\right) \; .
\end{equation}

Replacing the interface condition (\ref{interfacepsi}) in (\ref{ic}), it is possible to write
\begin{equation}
\psi_B'^+ = \alpha \psi_B'^- + \gamma \psi_B^- \; ,
\end{equation}
where
\begin{equation}
\alpha = \left(\frac{4v_1^2(E- \Delta_2)+(v_2^2-v_1v_2)(E- \Delta_1)}{4v_2^2(E- \Delta_1)+(v_1^2-v_1v_2)(E- \Delta_2}\right)
\end{equation}
and 
\begin{equation}
\gamma = k_y(1-\beta)\left(\frac{v^2_2(E- \Delta_1)-v_1^2(E- \Delta_2)+v_1v_2(\Delta_2 -\Delta_1)}{4v^2_2(E- \Delta_1)+(v^2_1-v_1v_2)(E- \Delta_2)}\right) \; .
\end{equation}

Analogously,  for $\psi_A$, we find

\begin{equation}
\psi_A'^+=\zeta \psi_A'^- +\eta \psi_A^-
\end{equation}

and

\begin{equation}
\psi_A^+=\kappa \psi_A^- \; ,
\end{equation}
where
\begin{equation}
\zeta = \left(\frac{4v_1^2(E+ \Delta_2)+(v_2^2-v_1v_2)(E+ \Delta_1)}{4v_2^2(E+ \Delta_1)+(v_1^2-v_1v_2)(E+ \Delta_2}\right)\; ,
\end{equation}

\begin{equation}
\eta = k_y(1-\kappa)\left(\frac{v_1^2(E+\Delta_2)-v^2_2(E+ \Delta_1)+v_1v_2(\Delta_2 -\Delta_1)}{4v^2_2(E+ \Delta_1)+(v^2_1-v_1v_2)(E+ \Delta_2)}\right)
\end{equation}

and

\begin{equation}
\kappa = \left(\frac{8v_1^2(E+ \Delta_2)+(11v_2^2-v_1^2-2v_1v_2)(E+ \Delta_1)}{8v_2^2(E+ \Delta_1)+(11v_1^2-v_2^2-2v_1v_2)(E+ \Delta_2)}\right) \; .
\end{equation}
One can see that the interface conditions for $\psi_B$ are equal to the interface conditions for $\psi_A$ with the exchange of $E$ by $-E$ and $k_y$ by $-k_y$.

Thus, the boundary conditions for $\psi_B$ are given by
\begin{equation}
\psi_B \left(\frac{a}{2}^+\right)=\beta_1 \psi_B \left(\frac{a}{2}^-\right)\;,
\label{11}
\end{equation}
\begin{equation}
\psi_B'\left(\frac{a}{2}^+\right)=\alpha_1 \psi_B'\left(\frac{a}{2}^-\right) + \gamma_1 \psi_B \left(\frac{a}{2}^-\right) \;,
\label{21}
\end{equation}
\begin{equation}
\psi_B \left(\frac{a}{2}+b^+\right)=\beta_2 \psi_B \left(\frac{a}{2}+b^-\right)\;,
\label{31}
\end{equation}
\begin{equation}
\psi_B '\left(\frac{a}{2}+b^+\right)=\alpha_2 \psi_B '\left(\frac{a}{2}+b^-\right) \gamma_2 \psi_B \left(\frac{a}{2}+b^-\right) \;,
\label{41}
\end{equation}
\begin{equation}
\psi_B(a+b)=e^{iK(a+b)}\psi_B(0)
\label{bloc11}
\end{equation}
and
\begin{equation}
\psi_B '(a+b)=e^{iK(a+b)}\psi_B '(0)\;,
\label{bloc111}
\end{equation}
where $\beta_1$, $\alpha_1$ and $\gamma_1$ are equal to $\beta$, $\alpha$ and $\gamma$, respectively, with $\Delta_1 = \Delta$ and $\Delta_2 = 0$. Whereas, $\beta_2$, $\alpha_2$ and $\gamma_2$ are equal to $\beta$, $\alpha$ and $\gamma$, respectively, with $\Delta_1 = 0$, $\Delta_2 = \Delta$ and with the exchange of $v_1$ by $v_2$ and vice-versa.

\subsection{Superlattice minibands}

The general solution for  Eq.(\ref{uncoupled}) is 

{\small
\begin{equation}
 \psi_B=\left\{ \begin{array}{cc}

                   A\cos k_1x+B\sin k_1x, \ \ \ \  & 0\leq x<\frac{a}{2}  \\
                   C\cos k_2x+D\sin k_2x, \ \ \ \  & \frac{a}{2}\leq x<\frac{a}{2}+b \\
                   E\cos k_1x+F\sin k_1x, \ \ \ \  & \frac{a}{2}+b\leq x\leq a+b 
        
\end{array}
 \right.\;,\label{pp}
\end{equation} } 
where $A$, $B$, $C$ $D$, $E$ and $F$ are constant coefficients, $k_1^2=\frac{E^2-\Delta^2}{\hbar^2 v_1^2}-k_y^2$ and $k_2^2=\frac{E^2}{\hbar^2 v_2^2}-k_y^2$.

The derivative of Eq.  (\ref{pp}) is given by

{\small
\begin{equation}
\psi'_B=\left\{
\begin{array}{cc}
             -k_1A\sin k_1x+q_1B\cos k_1x, & 0\leq x<\frac{a}{2}  \\
             -k_2C\sin k_2x+q_2D\cos k_2x, & \frac{a}{2}\leq x<\frac{a}{2}+b  \\ 
             -k_1E\sin k_1x+q_1F\cos k_1x, & \frac{a}{2}+b\leq x\leq a+b  
       
\end{array}
 \right.\;.\label{qq}
\end{equation}} 

Replacing the wave function (\ref{pp}) and its derivative (\ref{qq}) in the boundary conditions (\ref{11}) - (\ref{bloc111}) we get a system of six equations with six coefficients. Solving them as was done in \cite{jrflima}, we obtain the relation

\begin{eqnarray}
\cos k_x (a+b)&=&\cos k_1a\cos k_2 b  \nonumber \\
 &-&\frac{1}{2}\left[\beta_2 \alpha_1 \frac{k_1}{k_2}+\beta_1 \alpha_2 \frac{k_2}{k_1}-\frac{\gamma_1 \gamma_2}{k_1 k_2}\right]\sin k_1a\sin k_2b \nonumber \\
 &+&\frac{1}{2}\left[\frac{\gamma_1 \alpha_2 +\beta_1 \gamma_2}{k_1}\right]\sin k_1a \cos k_2 b \nonumber \\
 &+&\frac{1}{2}\left[\frac{\alpha_1 \gamma_2 + \gamma_1 \beta_2}{k_2}\right]\cos k_1 a \sin k_2 b \; ,
\label{band}
\end{eqnarray}
which is our main result. In this equation,  $k_x$, the $x$ component of the carrier wavevector, replaces the Bloch wavenumber $K$, since in this problem they coincide.

\begin{figure}[h!]
\centering
\includegraphics[width=9cm,height=4cm]{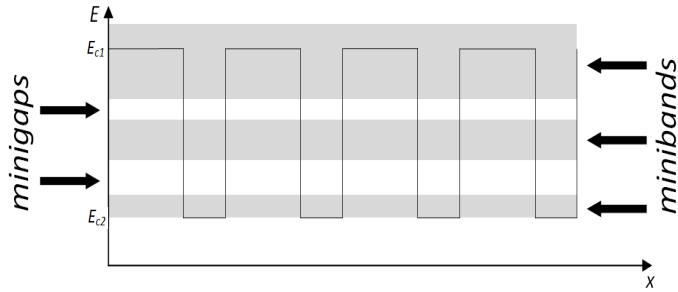}	
\caption{The conduction band of the graphene sheet as a function of the position $x$ showing the induced minibands (gray) and minigaps (white)  between the lower edges, $E_{c1}$ and $E_{c2}$, of the conduction band in each region of influence of the substrate.}\label{miniband}
\end{figure}

The left hand side of Eq.(\ref{band}) is limited in the interval (-1,1). For this reason, in the right hand side one has forbidden values for the energy. This implies in the appearance  of \textit{energy bands} with \textit{gaps} where initially was  the conduction  band. In summary, the heterostructured  substrate  induces a variable bandgap  in graphene, as shown in Fig. \ref{estrutura}. The periodicity of the induced properties of graphene (bandgap width and Fermi velocity) leads to the further opening of minigaps between the bottoms $E_{c1}$ and $E_{c2}$ of the conduction band in the differently gapped regions of the graphene sheet, as shown in Fig. \ref{miniband}.

Solving for $\psi_A$ we get the same relation (\ref{band}) with different coefficients which come from the interface conditions for $\psi_A$. As mentioned previously, the boundary conditions for $\psi_B$ are equal to the boundary conditions for $\psi_A$ with the simultaneous exchange of $E$ by $-E$ and $k_y$ by $-k_y$. So, the spectrum $E(k_y)$ is symmetric under  double reflection over both axes.  This leads to an indirect band gap. When the Fermi velocity is constant, $\beta =1$, which makes $\gamma =0$. In this case, $k_y$ appears in the spectrum only squared in $k_1$ and $k_2$, so the spectrum is independently invariant under the transformation $k_y \rightarrow -k_y$, making the band gap direct.

\begin{figure}[h]
\centering
\includegraphics[width=10cm,height=5cm]{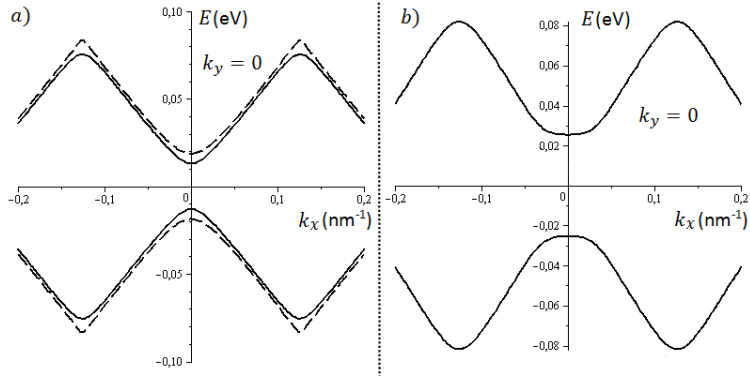}	
\caption{The dispersion relation obtained from (\ref{band}) with $a = b = 12.5$nm. In a) we consider $v_1 = v_2 = 10^6$m/s to compare our result with reference \citep{ratnikov}. The dashed curves are plots of Eq. (12) from reference \citep{ratnikov}. In b) one has $v_1 = 1.49\cdot 10^6$m/s and $v_2 = 10^6$m/s.}\label{comparacao}
\end{figure}

In Fig. \ref{comparacao}a we compare our result for the dispersion relation (\ref{band}) with that of reference  \citep{ratnikov} for $k_y=0$, obtained using the transfer matrix method, for the same Hamiltonian. In order to do that, we made $v_F= const.$  and used the same values of the lattice parameters $a$ and $b$ as in \citep{ratnikov} . As seen in the figure, the agreement is excellent.  Now, let us see the effect of considering the spatial variation of the Fermi velocity:  Fig. \ref{comparacao}b, again with $k_y=0$, as compared to Fig. \ref{comparacao}a shows a widening of the bandgap plus a slight flattening of the bands in the gap region.

\begin{figure}[h!]
\centering
\includegraphics[width=10cm,height=5cm]{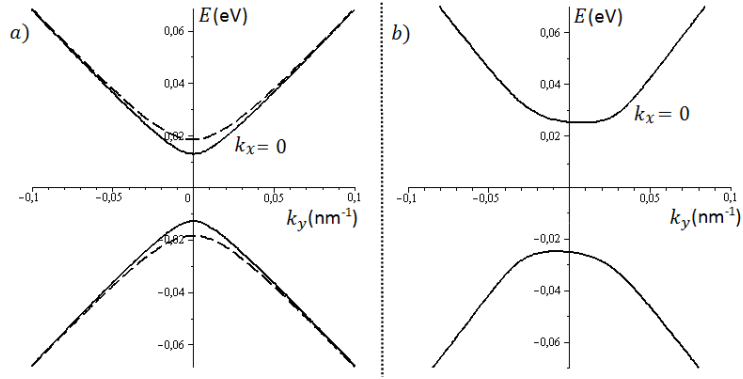}	
\caption{The energy in terms of $k_y$ with $k_x=0$ and $a = b = 12.5$nm. In a) we consider $v_1 = v_2 = 10^6$m/s to compare our result with reference \citep{ratnikov}. The dashed curves are plots of Eq. (12) from reference \citep{ratnikov}. In b) one has $v_1 = 1.49\cdot 10^6$m/s and $v_2 = 10^6$m/s.}\label{evsky}
\end{figure}

In Fig. \ref{evsky}a we plot the energy in terms of $k_y$ with $k_x=0$. Again, to compare with reference \citep{ratnikov} the Fermi velocity was fixed. Once again we get excellent agreement.  Fig. \ref{evsky}b shows the case of varying Fermi velocity. Not only the gap widened but it suffered a direct to indirect transition. Here, it is clear  the symmetry of the bands under reflection over both the $E$ and the $k_y$ axes.

\section{Conclusions}

We verified that modulating both the band gap and Fermi velocity in a single layer graphene sheet on a periodic substrate leads to the opening of an indirect band gap for the resulting graphene superlattice. Since a gap is of fundamental importance for electronic applications, band and Fermi velocity engineering leading to the control of both the type and width of the gap become important tools for the design of graphene-based electronic devices. We studied the specific case of graphene on alternating SiO$_2$ and \textit{h}-BN, but the model proposed here can be used in the case of a generic substrate composition. Our result was compared with the known results obtained via the transfer matrix method for  constant Fermi velocity with excellent agreement. 

{\bf Acknowledgements}: This work was partially supported by  CNPq, CNPQ-MICINN binational,  CAPES-WEIZMANN binational and CAPES-NANOBIOTEC.

\bibliographystyle{elsarticle-num} 
\bibliography{ref}

\end{document}